# SUBJECTIVE EVALUATION OF FORMS IN AN IMMERSIVE ENVIRONMENT

Jean-François Petiot, Damien Chablat

Ecole Centrale de Nantes
IRCCyN – UMR CNRS 6597 – Equipe MCM
1, rue de la Noë BP 92101
44321 NANTES Cedex 3 France
Phone : 33 2 40 37 69 59 – Fax : 33 2 40 37 69 30
E-mail : {Jean-Francois.Petiot, Damien.Chablat}@irccyn.ec-nantes.fr

**Abstract :** User's perception of product, by essence subjective, is a major topic in marketing and industrial design. Many methods, based on users' tests, are used so as to characterise this perception. Methods like multidimensional scaling, semantic differential method, and preference mapping are well known in sensorial analysis or in the food industry. These methods are used in order to built a perceptual space in order to position a new product, to specify requirements by the study of user's preferences, to evaluate some product attributes, related in particular to style (aesthetic). These early stages of the design are essential for a good orientation of the project. In parallel, virtual reality tools and interfaces are more and more efficient for suggesting to the user complex feelings, and creating in this way various levels of perceptions. In this article, we present on an example an adaptation of multidimensional scaling and preference mapping for the subjective assessment of virtual products. These products, which geometrical form is variable, are defined with a CAD model and are proposed to the user with a spacemouse and stereoscopic glasses. Advantages, uses and limitations of such evaluation is next discussed.

## 1- Introduction

In today's highly competitive market, developing new products that meet consumers' tastes is a crucial issue in product design. To improve attractiveness, a well-designed product should not only satisfy requirements, but should also satisfy consumers' psychological needs, by essence subjective. For many products, the aesthetic properties are as important as technical functions [1].

In order to predict the success of a product, to control and to optimise its performances, it's time now to take into account esteem and aesthetics functions in the beginning of the design. Although industrial and product designers are keenly aware of the importance of design aesthetics [2], there is an obvious lack of systematic, scientific, and engineering methods to help them make aesthetic design decisions and conduct aesthetic evaluation. For example, the perception of the shape of a product is often nothing but a style of design, depending much more on the designer's taste than on real customers' trends, as some studies clearly showed [3].

In this context, Virtual Reality (VR) seems to offer promising functionalities for the assessment of virtual products [4]. The Virtual reality interfaces available on the market are now mature enough for suggesting to the user relevant feelings and sensations [5]. The main problem is now to learn how to use it and to define relevant methods for their integration into the design process. How do we use virtual products to study the aesthetics of products and to help make them more aesthetic ? How increase the attractiveness of the shape of a product via virtual reality tools ?

In this article, we present the use of VR tools and user-tests for the study of aesthetics. It is assumed that we possess an innate ability to perceive a wide range of qualities in products that shape our response to them [6]. But it's often more difficult to detect bad proportions in the design of products, or to explain rules which lead to a feeling of harmony and beauty. A study based on the golden section or "divine proportion", described in [7], shows that it is not always clear whether the artists consciously used the golden section or whether they intuitively approximated the associated ratio.

Aesthetics is clearly a multidimensional characteristic, not independent of ergonomics and usability considerations. The concept of ergo-aesthetics, proposed in [8], refer to the integrated design approach that is aimed at meeting both ergonomic and aesthetic design objectives. Two main approaches can be used for the study of aesthetics [8]:

- The top-down approach attempts to understand aesthetic responses as a whole,
- The bottom-up approach attempts to identify the basic pictorial features and compositional patterns that please or displease the senses.

In this article, we propose an illustration of both approaches for the evaluation of forms in an immersive environment. In section 2, we present two classical methods in sensorial analysis which can be used in product design for grasping the subjective response of subjects. In section 3, we propose an adaptation of these methods to the experimental study of the "appeal" of forms. The perception of the aesthetic of forms by a subject is carried out in a VR environment, and with particular products: table glasses. The main results of this experiment are discussed in section 4, and perspectives for the use of VR for the design of forms are discussed.

**2- Background**

To study users' perceptions, researchers in marketing and sensorial analysis propose various methods [9][10][11]. Perceptual maps are commonly used to take perceptions into account and to control the product positioning. The basic idea is to build a multi-attribute perceptual space in which each product is represented by a point. We propose to describe *multidimensional scaling* (MDS), a classical method for building the perceptual space, and *preference mapping* techniques, which are relevant to grasp preference assessments. An adaptation of both methods will be used in section 3.

*2.1 – MDS*

Multidimensional scaling uses dissimilarity assessments to create a geometrical representation of a family of "stimuli". This method, developed initially for psychometric analysis [12], is a process whereby a distance matrix among a set of stimuli is translated into a representation of these stimuli inside of a perceptual space. Taking all the possible pairs of stimuli (here pairs of products) into account, the subject evaluates their degree of similarity on a quantitative scale and fills in this way a dissimilarity matrix. Technically, what MDS does is to find a set of points $(X_i)_{i=1,...,N}$ in a K-dimensional space such that the distances among them correspond as closely as possible to the dissimilarity (or a function of it) in the input matrix, according to a given optimisation criterion. This criterion can be for example a function called *stress,* (equation 1) which represents the "badness of fit", and the distance can be the Euclidean distance (equation 2), but several different expressions can be used [13]. $D_{ij}$ is the perceptual dissimilarity between stimuli i and j, $d_{ij}$ the Euclidean distance, $x_{ik}$ the coordinate of stimuli *i* on dimension *k*, K the number of dimensions of the perceptual space, selected by the experimenter. After computation, each stimuli is represented in the Euclidean space by a point $X_i(x_{i1}, ..., x_{iK})$.

$$stress = \left[ \frac{\sum_i \sum_j (d_{ij} - D_{ij})^2}{\sum_i \sum_j d_{ij}^2} \right]^{1/2} \quad (1) \qquad d_{ij} = \sqrt{\sum_{k=1}^{K} (x_{ik} - x_{jk})^2} \quad (2)$$

The main advantage of this method is that the tests are based on instinctive dissimilarity assessments, which do not impose any criteria or predefined semantic scale. This method provides a space for a visualisation of the relevant dimensions for stimuli's perception. It is used in marketing and sensorial analysis, the stimuli can be sounds, fragrance, products, concepts,…It is well suited to study the relationship between products. We have used in the next section metric MDS, based on the minimisation of the stress, and using a sparse matrix of dissimilarity.

*2.1 – Preference mapping*

Preference mapping constitutes a group of statistical techniques which are also common in sensory analysis and marketing research. Internal (MDPREF) and external (PREFMAP) preference mapping both generate graphic interpretations of individual preferences [14][15]. To construct such preference mappings, multiple regression techniques are used. The principle is to find a correlation between product's preference and product's position in the perceptual space. Technically speaking, one must perform a multiple linear regression of the preference on the perceptual coordinates. These techniques need some strong assumptions. The preference's model is supposed to be linear most of the time, it's not possible to consider threshold phenomena for the preference, and the method needs an absolute evaluation of the preference as input (hedonic scale). The use of a hedonic scale to assess the preference in an absolute manner encounters some difficulties and limitations which can be avoided by the use of pairwise comparison methods [16]. In the PREFMAP model, consumers are represented by their ideal point in the perceptual space, which represents the most preferred combination of attributes. Other models with infinite preferred attribute levels are called 'vector models'. We use in the next section preference mapping based on the vector model.

**3- Experimental study**

*3.1 – Description of the VR tools*

The main objective of the experiment is to show that the study of the aesthetic properties of forms can be conducted on virtual products, and that an immersive environment is particularly helpful for the assessment and the improvement of aesthetics.

To illustrate our approach, we propose to study the aesthetics of table glasses, which are very interesting products from an aesthetics and esteem point of view. A study on such products (wine–glass) is proposed in [17], where the authors present a method for form generation, and in [18], where a methodology for a solid assessment of product semantics is proposed.

A digital model of glasses with various forms was designed with the CAD and VR software *Catia V5R11*. A texture with the material "glass" were applied to the models of glass. A Spacemouse and stereoscopic eyeglasses were used by the subject for enhancing his/her perception of the environment, and for facilitating the various assessments.

*3.2 – Description of the assessment tests*

Several forms of table glasses were generated with the CAD software. All glasses are intended to the same usage, for example wine glasses. They all have the same general form, balloon glass, and are made off three parts: base, foot and container. The mathematical model of the curve for the generation of forms is given figure 1. The base, the foot, and the container are modelled with a spline-curve with 15 control nodes. Continuity conditions are added at points P5 and P9. The inner shape of the container is parallel to the outer shape, with a gap of 1 mm. The 3D solid is generated by rotation of the curve.

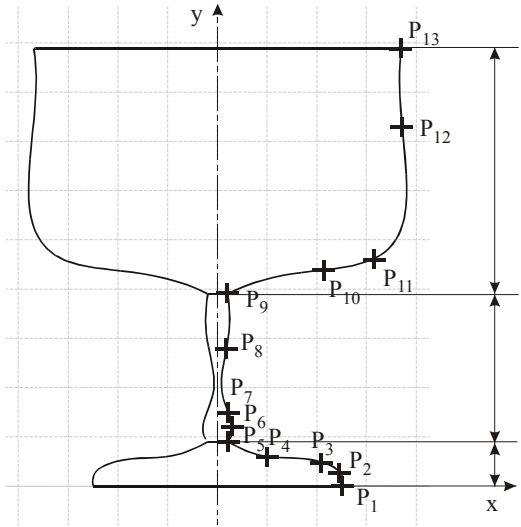

***Fig. 1*** : Model of the outer curve of the glass.

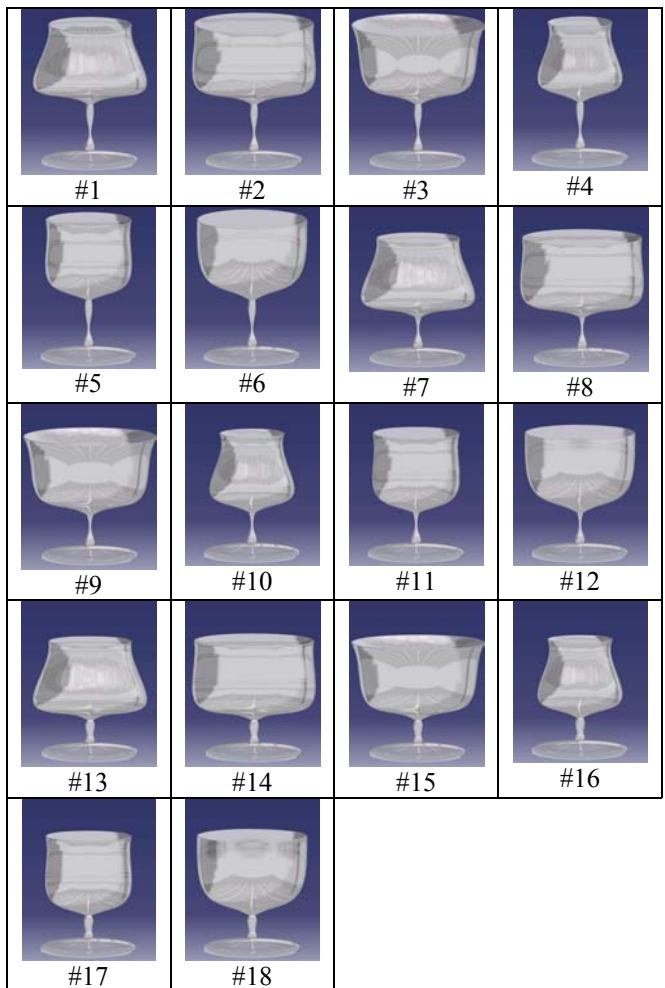

***Fig. 2*** : Pictures of the 18 glasses.

All the glasses used for the experimental study are an "instanciation" of this model. 18 glasses were generated (figure 2), which constitute the stimuli for the assessment tests.

VR tools were next used for the evaluation of the aesthetics qualities of the glasses The assessment performed by the subject was made of several stages, presented in the following sections.

### 3.2.1 –Stage 1: Building of the perceptual space

The aim of this stage is to find how many dimensions can describe the aesthetic, or the "appeal" of the glasses, and to find the location of the glasses according to these dimensions. For that, the subject had to quantify the perceived differences between pairs of products from an aesthetic point of view. This stage belongs to the "top-down" approach.

The subject was asked to say to what degree the products were different. He/she chose between "identical", "a little different", "different" and "very different". The answer was coded on an integer scale from 0 (identical) to 3 (very different). In order to reduce the assessment task, the subject was not forced to fill all the dissimilarity matrix. He only had to chose the more obvious comparisons. He had the complete choice of the comparisons to fill in the $N.(N-1)/2=153$ potential comparisons of the superior half of the dissimilarity matrix. Nevertheless, so as to have computable data, and to reveal the multidimensionality of aesthetics, each product should be involved in at least three comparisons.

The pairwise comparison were performed by the subject with the VR software and easy "click and drag" operations. The (sparse) dissimilarity matrix provided by the subject is given table 1.

|    | 1 | 2 | 3 | 4 | 5 | 6 | 7 | 8 | 9 | 10 | 11 | 12 | 13 | 14 | 15 | 16 | 17 | 18 |
|----|---|---|---|---|---|---|---|---|---|----|----|----|----|----|----|----|----|----|
| 1  | 0 | 2 | * | 1 | * | 1 | * | 0 | * | *  | 1  | *  | *  | *  | 2  | *  | *  | *  |
| 2  | 2 | 0 | 1 | 3 | 2 | 1 | 2 | 0 | 1 | 3  | 2  | 1  | 1  | 0  | 1  | 3  | 2  | 1  |
| 3  | * | 1 | 0 | 2 | 2 | * | * | * | 0 | 2  | 2  | *  | *  | 2  | 0  | *  | 3  | *  |
| 4  | 1 | 3 | 2 | 0 | 2 | 2 | 1 | 3 | 2 | 0  | 1  | 3  | 2  | 3  | 2  | 1  | 2  | 2  |
| 5  | * | 2 | 2 | 2 | 0 | * | 1 | * | * | *  | 0  | *  | *  | *  | 2  | *  | *  | *  |
| 6  | * | 1 | * | 2 | * | 0 | * | * | * | *  | *  | *  | *  | *  | 2  | *  | *  | *  |
| 7  | 0 | 2 | * | 1 | 1 | * | 0 | 2 | * | *  | 3  | 2  | 0  | 2  | 0  | *  | *  | *  |
| 8  | * | 0 | * | 3 | * | * | 2 | 0 | * | *  | 3  | *  | *  | *  | 1  | *  | *  | *  |
| 9  | * | 1 | 0 | 2 | * | * | * | * | 0 | *  | 3  | *  | *  | *  | 0  | *  | *  | *  |
| 10 | * | 3 | 2 | 0 | * | * | * | * | * | 0  | 1  | *  | *  | *  | 2  | *  | *  | *  |
| 11 | 1 | 2 | 2 | 1 | 0 | * | 3 | 3 | 3 | 1  | 0  | 1  | 2  | 3  | 3  | 1  | 0  | 3  |
| 12 | * | 1 | * | 3 | * | * | 2 | * | * | *  | 1  | 0  | *  | *  | 2  | *  | *  | *  |
| 13 | * | 1 | * | 2 | * | * | 0 | * | * | *  | 2  | *  | 0  | *  | 2  | *  | *  | *  |
| 14 | * | 0 | 2 | 3 | * | * | 2 | * | * | *  | 3  | *  | *  | 0  | 2  | *  | *  | *  |
| 15 | 2 | 1 | 0 | 2 | 2 | 2 | 0 | 1 | 0 | 2  | 3  | 2  | 2  | 2  | 0  | 2  | 1  | 1  |
| 16 | * | 3 | * | 1 | * | * | * | * | * | *  | 1  | *  | *  | *  | 2  | 0  | *  | *  |
| 17 | * | 2 | 3 | 2 | * | * | * | * | * | *  | 0  | *  | *  | *  | 1  | *  | 0  | *  |
| 18 | * | 1 | * | 2 | * | * | * | * | * | *  | 3  | *  | *  | *  | 1  | *  | *  | 0  |

***Tab. 1*** : Sparse dissimilarity matrix.

Next, we used MDS for determining how many dimensions can represent the dissimilarities, and for building a perceptual space which expresses these differences.

With this sparse matrix as input, an own implementation of metric MDS has been used to calculate the perceptual coordinates of the glasses [19]. A 2-dimensional configuration, with a stress value equal to 0.12 (considered as a correct "badness of fit") has been retained (figure 3).

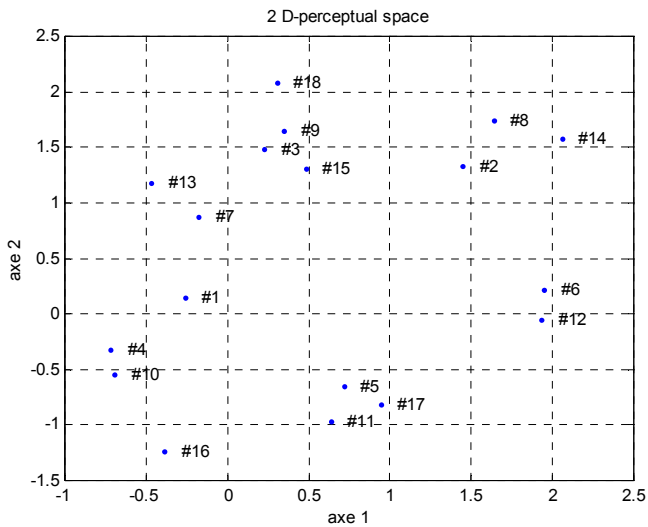

*Fig. 3* : 2D Perceptual space of the glasses

These results reveal the two main dimensions (axe1, axe2) which are relevant for describing the differences between the products from an "appeal" point of view.

*3.2.2 –Stage 2: assessment of the appeal*

The aim of this stage is to get an absolute assessment of the appeal, on a continuous scale. The task is the same as previous stage, but the subject is now forced to provide a mono-dimensional answer. In order to reveal the multidimensionality of aesthetic during stage 1, it's important that the subject do not carry out stage 1 from the results of stage 2. In other words, stage 1 must be performed before stage 2.

The subject was asked to rank the products in the order of increasing appeal, or into piles of similar appeal. The subject associated to each product (or each pile) an absolute value of the appeal, on a continuous scale from 0 to 10. This scale is similar to the "hedonic" evaluation used in the food industry for preference assessments. The assessment of the glasses, provided by the subject, is given table 2.

| Appeal P | 0 | 1 | 2 | 3 | 4 | 5 | 6 | 7 | 8 | 9 | 10 |
|---|---|---|---|---|---|---|---|---|---|---|---|
| Glass # | 2 | 8 14 | 12 18 | 9 15 | 6 3 | 17 | 13 7 | 1 | 11 5 | 16 | 4 10 |

*Tab. 2* : Appeal scores of the glasses.

*3.2.3 –Stage 3: a model of appeal*

The main objective is to find how sensitive is the subject in detecting small variations in aesthetics, and to find his abilities to perceive and judge values, changes, and variations in design parameters. This stage belongs now to the "bottom-up" approach [20].

For that, three shape regulating rules were proposed to the subject. These rules correspond to a modification of the three determining design parameters of a glass, proposed in [21]. They are expressed as following (figure 4):

- R1: increase the total height of the container, d1,
- R2: increase the height of the foot, d2,
- R3: increase the diameter of the container, d3.

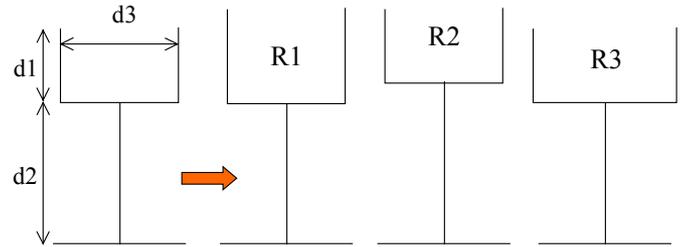

*Fig. 4* : Description of the shape regulating rules

For each glass Gj, the subject has to indicate the effect of the proposed shape regulating rule on his/her assessment of the appeal:

- If rule "Ri" increases the appeal, ΔPi(Gj) = 1
- If rule "Ri" does not change the appeal, ΔPi(Gj) = 0
- If rule "Ri" decreases the appeal, ΔPi(Gj) = -1

This is equivalent to an assessment of the "derivative" of the appeal according to the design parameters.

The virtual reality model of the glass is particularly suitable for the required assessments. Indeed, the subject can even make the modification on the virtual model to visualise and compare the differences, if the assessment of the appeal change is not obvious.

The values of the dimensions d1, d2, d3 for each glass and the assessments proposed by the subject are given in the following table 3. For each shape regulating rule, these values ΔPi(Gj) are proportional to the partial derivative $\frac{\partial P}{\partial dj}$ , with a positive multiplicative factor kj.

| Glass # | d1 cm | d2 cm | d3 cm | R1 $k1\frac{\partial P}{\partial d1}$ | R2 $k2\frac{\partial P}{\partial d2}$ | R3 $k3\frac{\partial P}{\partial d3}$ |
|---|---|---|---|---|---|---|
| G1 | 8 | 7 | 8 | -1 | 0 | -1 |
| G2 | 8 | 7 | 9,5 | -1 | 1 | -1 |
| G3 | 8 | 7 | 9,5 | -1 | 0 | -1 |
| G4 | 8 | 7 | 6 | -1 | -1 | 0 |
| G5 | 8 | 7 | 7 | -1 | -1 | -1 |
| G6 | 8 | 7 | 9 | -1 | 0 | -1 |
| G7 | 8 | 5 | 8 | -1 | 1 | -1 |
| G8 | 8 | 5 | 9,5 | -1 | 1 | -1 |
| G9 | 8 | 5 | 9,5 | -1 | 1 | -1 |
| G10 | 8 | 5 | 6 | -1 | 1 | 0 |
| G11 | 8 | 5 | 7 | -1 | 1 | -1 |
| G12 | 8 | 5 | 9 | -1 | 1 | -1 |
| G13 | 8 | 3 | 8 | -1 | 0 | -1 |
| G14 | 8 | 3 | 9,5 | -1 | 1 | -1 |
| G15 | 8 | 3 | 9,5 | -1 | 1 | -1 |
| G16 | 8 | 3 | 6 | -1 | 1 | -1 |
| G17 | 8 | 3 | 7 | -1 | 1 | -1 |
| G18 | 8 | 3 | 9 | -1 | 1 | -1 |

*Tab. 3* : Dimensions of the glasses and assessment of the appeal changes according to the shape regulating rules.

### 3.2.3 – Building of an model of appeal

A quadratic model of the appeal P is proposed. The predicted value, denoted $\overline{P}$, is given by:

$$\overline{P} = a1.d1 + a2.d2 + a3.d3 + a4 d1^2 + a5.d2^2 + a6.d3^2 \\ + a7.d1.d2 + a8.d1.d3 + a9.d2.d3 + a10 \quad (3)$$

The predicted partial derivatives of P are given by:

$$\frac{\partial \overline{P}}{\partial d1} = a1 + (2.a4 + a7.d2 + a8.d3).d1$$
$$\frac{\partial \overline{P}}{\partial d2} = a2 + (2.a5 + a7.d1 + a9.d3).d2 \quad (4)$$
$$\frac{\partial \overline{P}}{\partial d3} = a3 + (2.a6 + a8.d1 + a9.d2).d3$$

The set of coefficients {a1, a2,…, a10} is determined with an optimisation procedure, similar to the least square regression. It's a generalisation of the linear regression of the appeal on the design parameters, taking into account more data concerning the partial derivative.

With all observations $(P_i, \frac{\partial P_i}{\partial d j})$, the vector of variables $X$ = (a1, a2,…, a10, k1, k2, k3) is determined by minimising the following function $F$:

$$\min F(X) = \sum_{i=1}^{N}(P_i - \overline{P_i})^2 + \sum_{i=1}^{N}\sum_{j=1}^{3}(\frac{\partial P_i}{\partial d_j} - k_j \frac{\overline{\partial P_i}}{\partial d_j})^2 \quad (5)$$

$$w.r.t. X = (a_1, a_2, ..., a_{10}, k_1, k_2, k_3)$$

$P_i$ is the appeal, $\frac{\partial P_i}{\partial dj}$ the partial derivative, provided by the subject, for glass $G_i$,

$\overline{P}_i$ is the appeal, $\frac{\partial \overline{P}_i}{\partial dj}$ the partial derivative, provided by the model, for glass $G_i$,

The result of the optimisation procedure is given table 2.

| a1 | a2 | a3 | a4 | a5 | a6 | a7 | a8 | a9 | a10 |
|---|---|---|---|---|---|---|---|---|---|
| 3.6 | 0.12 | 13.6 | -1.52 | 0.02 | -0.13 | -0.01 | -1.71 | 0.01 | 82.36 |

| k1 | k2 | k3 |
|---|---|---|
| -0.01 | 14.15 | 0.01 |

**Tab. 4** : Coefficients of the model of appeal.

The response surface, corresponding to this model, is plotted in the next section.

## 4- Results and discussions

### 4.1 – Preference mapping 1

The first analysis which can be done is to try to explain the appeal of the glasses by the perceptual dimensions given by MDS.

Like for the PREFMAP model, this is done by a multiple regression using the perceptual axes (axe1, axe2) as independent variables and the appeal P as the dependent variable (equation 4).

$$P = a.axe\,1 + b.axe\,2 + c \quad (6)$$

The results of the regression is given table 5.

| a | b | c | $R^2$ | F | Significant[1] (p=0.01) |
|---|---|---|---|---|---|
| -2 | -1,8 | 7 | 0,91 | 80 | yes |

**Tab. 5** : Coefficients of the model of preference mapping 1

The regression is significant and the determination coefficient $R^2$ is close to 1. So the linear model is well adapted for describing the data provided by the subject. This good correlation shows a good coherency of the subject: the data provided in stage 1 are quite compatible with the data provided in stage 2.

Furthermore, the VR interfaces proposed to the subject do not introduce incoherence or "noise" in the assessment of the shape. In other words, the good coherence of the data is a sign for the perfect adaptation of the interface to the task proposed to the subject.

So as to give a graphical interpretation of the regression, the vector model of the appeal is plotted in the perceptual space. The origin of the vector is located arbitrarily in the origin of the frame, the values of the regression coefficients (a, b) give the orientation of the arrow, the arrowhead points in the direction of increasing appeal (figure 5). It can be shown that this vector is parallel to the steepest slope line of the plane (equation (6)). The perpendiculars to the vector are the iso-appeal curves.

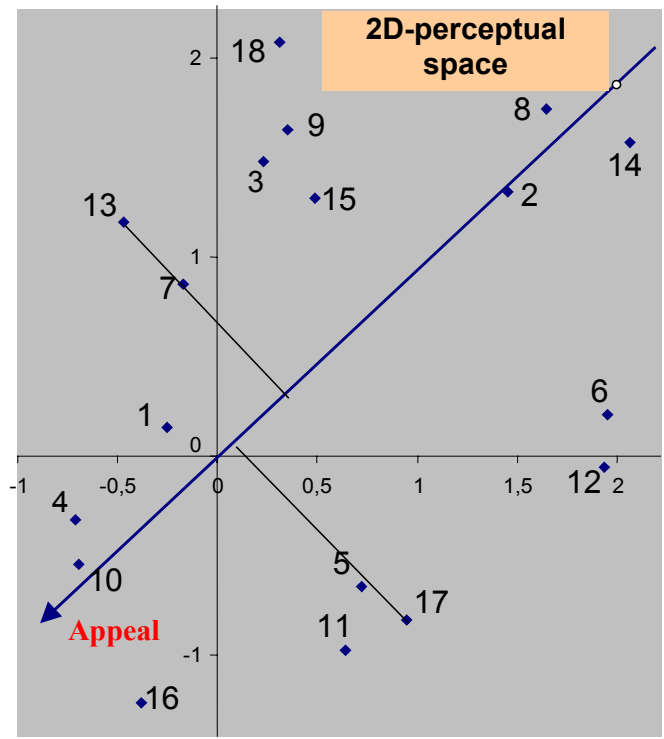

**Fig. 5** : vector model of appeal in the perceptual space

---
[1] according to a Fisher-Snedecor table

The interest of this graph lies in its ability to study the multidimensionality of aesthetics. What differentiate glass #13 and #17, which have about the same appeal ? May a particular attribute, or a particular design parameter, explain the difference ? Further studies are needed for answering these questions.

*4.2 – Preference mapping 2*

The second analysis which can be done is to try to explain the appeal of the glasses by certain design parameters of the forms.

The response surface of the appeal $\overline{P}$, obtained in section 3, can be plotted, according to the design variables (d2, d3) (equation 7)[2]:

$$\overline{P}=8.a1+a10+64.a4+(a2+8.a7).\mathbf{d2}+(a3+8.a8).\mathbf{d3}$$
$$+a5.\mathbf{d2}^2+a6.\mathbf{d3}^2+a9.\mathbf{d2}.\mathbf{d3} \quad (7)$$

The colormap of this surface is presented figure 6.

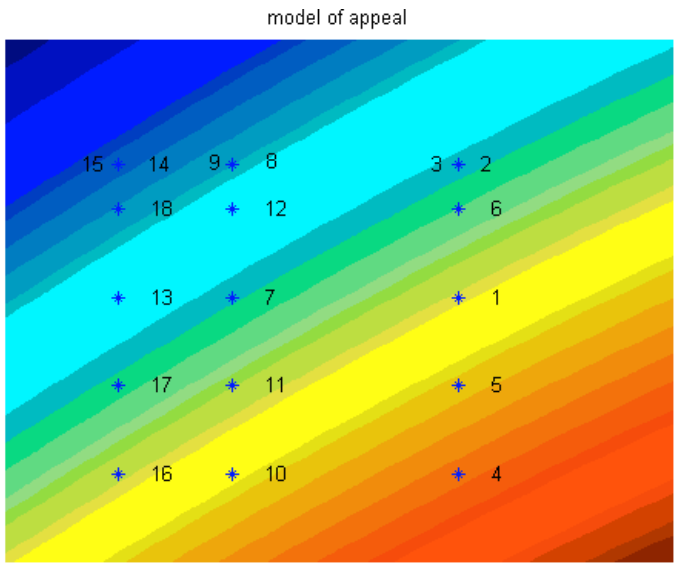

**Fig. 6**  Colormap of the appeal in the design parameter space (d2, d3).

The structure of the appeal seems to be very simple and uni-dimensional. Increasing d2 and decreasing d3 lead to an increase of the appeal. The perpendicular to the steepest descent line, given by the relation d3=d2/2+c, is about an iso-appeal line (figure 7).

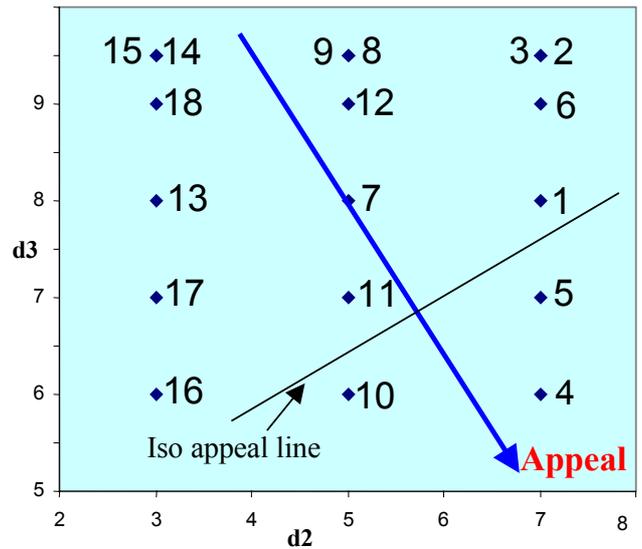

**Fig. 7** : Iso-appeal line in the design parameter space

The relation d3=d2/2+c is important for the study of the aesthetic of a glass. It's a characteristic of the subject, who has provided data which satisfy this relation. The strong point of our study is that the methodology presented allows the discovery of such relation. The subject is in most of the cases unconscious of this relation and is not able to formulate it explicitly.

Both preference mappings are interesting for improving the design of forms, or seeking "good" products because they use two complementary approaches.

**5- Conclusions**

We have presented in this paper an experimental study of the aesthetic of table glasses. This study of the "appeal" of the shapes has been carried out on virtual models, generated by a CAD/VR software, and with classical VR interface: a spacemouse and stereoscopic eyeglasses. With this tools, the study was based on the assessment by a subject of various characteristics of the glass. Two ways of investigations have been used: (1) a top-down approach, using multidimensional scaling and linear regression of the "appeal" of the shapes ; (2) a bottom-up approach, with the building of a model of the appeal, based on the assessment of "shape regulating rules ". A new method for establishing the model of appeal, based on an optimisation process, has been proposed.

Firstly, the good coherence of the data proposed by the subject shows the perfect adaptation of the interfaces to the proposed task. Furthermore, the subject does not encounter difficulties in the assessment tasks.

Secondly, the model of appeal, based on two design parameters of the shape of the glass, reveals how these parameters are linked for the subject's assessments.

In perspective, we will conduct a similar study using more design parameters, and based on principal component analysis. The next step will be to use a model of subjective evaluation for forms generation and product synthesis, approach used for example in Kansei engineering type 3

---

[2] for all glasses, the dimension d1=8cm

[22].

In future works, haptic devices will be introduced for the assessment of weight, stability and more generally ergonomic attributes.

**5- Acknowledgements**

We acknowledge the help of Hélène Compain for performing the tests and for various assessments of aesthetics.

**6- Bibliography**